\documentclass[12pt]{article}
\input epsf

\def\qqbar{\mbox{${\rm q\bar{q}}$}}
\def\ppbar{\mbox{${\rm p\bar{p}}$}}
\def\bbbar{\mbox{${\rm b\bar{b}}$}}

\newcommand{\msbot}{\ensuremath{m_{\tilde{\rm b}}}}
\newcommand{\mZ}{\ensuremath{m_{\rm Z}}}

\newcommand{\gevcc}{\ensuremath{{\rm GeV}\!/c^2}}

\newcommand{\epem}{\ensuremath{{{\rm e}^+{\rm e}^-}}}
\newcommand{\epemto}{\ensuremath{{{\rm e}^+{\rm e}^- \to}}}

\newcommand{\sbb}{\ensuremath{{\tilde{\rm b}}}}
\newcommand{\sq}{\ensuremath{{\tilde{\rm q}}}}

\newcommand{\sbsbbar}{\ensuremath{\sbb\bar{\sbb}}}
\newcommand{\sqsqbar}{\ensuremath{\sq\bar{\sq}}}

\newcommand{\eNPh}{\ensuremath{\varepsilon_{\rm NP}^{\rm had}}}

\def\Journal#1#2#3#4{{#1} {\bf #2} (#3) #4}

\def\PLB{{\em Phys. Lett.} {\bf  B}}

\def\PRL{\em Phys. Rev. Lett.}

\def\PRD{{\em Phys. Rev.} {\bf D}}
\def\ZPC{{\em Z. Phys.} {\bf C}}
\def\EPJ{{\em Eur. Phys. J.} {\bf C}}
\def\PREP{\em Phys. Rep.}
\def\CPC{\em Comput. Phys. Commun.}

\def\JPG{\em J. Phys. {\bf G}: Nucl. Part. Phys.}

\def\citedash{\hbox{--}\penalty\@m }

\begin{document}

\pagenumbering{arabic}
\pagestyle{plain}

\date{}
\title{ \null\vspace{1cm}
Closing the light sbottom mass window\\
from a compilation of \epemto\ hadron data 
\vspace{1cm}}
\author{Patrick Janot\\
\footnotesize CERN, PH Department, CH-1211 Geneva 23, Switzerland\\
\footnotesize e-mail: Patrick.Janot@cern.ch}

\maketitle

\begin{picture}(160,1)
\put(-5,110){\rm ORGANISATION EUROP\'EENNE POUR LA RECHERCHE NUCL\'EAIRE (CERN)}
\put(25,105){\rm Laboratoire Europ\'een pour la Physique des Particules}
\put(115,79){\parbox[t]{50mm}{\tt CERN-PH-EP/2004-006}}
%%\put(115,73){\parbox[t]{50mm}{\tt PHYSIC/2003-001}}
\put(115,73){\parbox[t]{50mm}{\tt 12-March-2004}}
\end{picture}

\vspace{.2cm}
\begin{abstract}
\vspace{.2cm}
The \epemto\ hadron cross section data from PEP, PETRA, TRISTAN, SLC and LEP, 
at centre-of-mass energies between 20 to 209\,GeV, are analysed to search
for the production of a pair of light sbottoms decaying hadronically via
R-parity-violating couplings. This analysis allows the 95\%\,C.L. exclusion 
of such a particle if its mass is below 7.5\,\gevcc. The light sbottom mass 
window is closed.
\end{abstract}

\vfill
\centerline{\it To be published in Physics Letters B}
\vskip .5cm
%\noindent
%--------------------------------------------\hfil\break
%{\ninerm $^*)$ See next pages for the list of authors}

\eject

\section{Introduction}

At the end of the last millenium, the Tevatron Collaborations~\cite{d0,cdf} 
came out with a bottom quark production cross section at $\sqrt{s}=1.8$\,TeV 
in excess of the theoretical prediction by about a factor of two. 
Refined parton density functions and other theoretical improvements, 
{\it e.g.}, in the b-quark fragmentation function, have 
recently been shown to account for the difference in the data 
recorded at $\sqrt{s}=1.96$\,TeV~\cite{mangano}. 

A more exotic model~\cite{berger}, in which a pair of gluinos with mass 12 to 
16\,\gevcc\ is produced in \ppbar\ collisions, with subsequent decays 
into a bottom quark and a light sbottom, with mass below 6\,\gevcc, has  
been shown to also fit the excess well.
In this model, the sbottom must either be long-lived or decay {\it via}
R-parity-violating coupling to light quarks, {\it e.g.}, $\tilde{\rm b} 
\to \bar{\rm u} \bar{\rm s}$, to comply with various experimental 
constraints. Long-lived sbottoms have recently been excluded 
up to masses of 92\,\gevcc\ by ALEPH~\cite{aafke} in direct searches 
for $\epemto\ \qqbar\sqsqbar$ and $\epemto\ \sqsqbar$, but 
R-parity-violating prompt hadronic decays have not been 
addressed by the ALEPH analysis.

A light, hadronically decaying sbottom would increase
the \epemto\ hadron cross section above the \sbsbbar\ production threshold
by up to a quarter of the \epemto\ \bbbar\ cross section, {\it i.e.,} about 
2\% far from the Z peak and 5\% at the Z peak. For this reason, 
the measurements of the hadronic cross section at centre-of-mass 
energies from 20 to 209\,GeV ({\it i.e.}, well above the 
known \bbbar\ resonances) from PEP~\cite{markii,mac}, 
%PETRA~{\catcode`@=11 [\hbox{\b@cello}--\hbox{\b@pluto}]}, 
%TRISTAN~{\catcode`@=11 [\hbox{\b@amy}--\hbox{\b@venusc}]} and 
PETRA~\cite{cello}--\cite{pluto}, TRISTAN~\cite{amy}--\cite{venusc} and 
LEP/SLC~\cite{lep}, are re-analysed in this letter to search 
for a possible consistent excess.

This letter is organized as follows. A compilation of the data 
is presented in a synthetic manner in Section~\ref{sec:data}
to allow easy re-interpretation in the future. The global fit 
of the data is described in Section~\ref{sec:fits}. The 
results of the analysis are given in Section~\ref{sec:results} and
the conclusions are listed in Section~\ref{sec:conclusion}.

\section{The hadronic cross section data}
\label{sec:data}

\subsection{The data from PEP, PETRA and TRISTAN}
\label{sec:lowdata}

Most of the data from PEP, PETRAN and TRISTAN are published 
under the form of the ratio $R$ of the effective Born hadronic cross section 
$\sigma^0_{\rm had}$ to the point-like $\epemto\ \mu^+\mu^-$ cross section
$\sigma^0_{\mu\mu}$, 
\begin{equation}
\sigma^0_{\mu\mu}(s) = {\alpha^2_{\rm QED}(s) \over \alpha^2_{\rm QED}(0)} \times
{86.85\,{\rm nb} \over s},
\end{equation}
where $s$ is the \epem\ centre-of-mass energy squared and $\alpha_{\rm QED}$
is the fine structure constant.
The latest TOPAZ~\cite{topazb,topazc} and VENUS~\cite{venusc} publications report 
directly the value of $\sigma^0_{\rm had}$ instead. In both cases, the latter 
includes a correction that unfolds the effects of initial state 
radiation (ISR), while still reflecting the running of the fine structure 
constant with the centre-of-mass energy~\cite{zenin}.

\begin{table}[htbp]
\begin{center}
\caption{\footnotesize The ratio $R$ and the effective Born hadronic cross 
section, $\sigma^0_{\rm had}$, from the PEP and PETRA experiments,
with increasing centre-of-mass energy ($\sqrt{s}$). The expected statistical 
($\sigma_{\rm stat}$), point-to-point systematic ($\sigma_{\rm ptp}$) and 
normalization systematic ($\Delta_{\rm norm}$) uncertainties are 
also given (in \%). The latter is correlated between all energy points 
in a given publication. An additional normalization error $\Delta_{\rm QED} 
= 0.1\%$, fully correlated between all measurements, is to be added to 
account for missing QED higher orders. The last column points to the original 
publication.
\label{tab:petra}}
\vspace{5mm}
{\scriptsize
\begin{tabular}{|c|c|c|ccc|l|} \hline\hline
\rule{0pt}{5.0mm}
$\sqrt{s}$ (GeV) & Ratio $R$ & $\sigma^0_{\rm had}$ (pb) & 
$\sigma_{\rm stat}$ (\%) & 
$\sigma_{\rm ptp}$ (\%) & 
$\Delta_{\rm norm}$ (\%) & 
Reference \\ 
\hline\hline
21.990 &  3.550 & 697.0 &   2.4 &   3.0 &   1.6 & MARKJ~\cite{markj}   \\
22.000 &  3.860 & 757.2 &   3.0 &   2.1 &   1.7 & CELLO~\cite{cello}   \\
21.990 &  3.860 & 757.9 &   2.3 &   0.0 &   3.5 & TASSO~\cite{tassob}  \\
22.000 &  4.110 & 806.2 &   3.1 &   0.0 &   2.4 & JADE~\cite{jade}     \\
22.000 &  3.470 & 680.7 &  18.3 &   0.0 &   6.0 & PLUTO~\cite{pluto}   \\
\hline
25.000 &  3.720 & 566.7 &  10.4 &   0.0 &   3.5 & TASSO~\cite{tassoa}  \\
25.000 &  4.030 & 613.9 &   5.1 &   3.0 &   1.6 & MARKJ~\cite{markj}   \\
25.010 &  4.240 & 645.4 &   6.5 &   0.0 &   2.4 & JADE~\cite{jade}     \\
\hline
27.500 &  3.910 & 493.3 &   8.2 &   0.0 &   3.5 & TASSO~\cite{tassoa}  \\
27.600 &  4.070 & 509.8 &   7.0 &   0.0 &   6.0 & PLUTO~\cite{pluto}   \\
27.660 &  3.850 & 480.2 &  12.5 &   0.0 &   2.4 & JADE~\cite{jade}     \\
29.000 &  3.920 & 445.3 &   1.3 &   0.0 &   2.3 & MARKII~\cite{markii} \\
29.000 &  3.960 & 449.8 &   0.8 &   0.0 &   2.3 & MAC~\cite{mac}       \\
\hline
29.930 &  3.550 & 378.8 &  11.8 &   0.0 &   2.4 & JADE~\cite{jade}     \\
30.100 &  3.940 & 415.8 &   4.5 &   0.0 &   3.5 & TASSO~\cite{tassoa}  \\
30.380 &  3.850 & 398.9 &   5.0 &   0.0 &   2.4 & JADE~\cite{jade}     \\
30.610 &  4.150 & 423.6 &   3.5 &   3.0 &   1.6 & MARKJ~\cite{markj}   \\
30.800 &  4.100 & 413.4 &   3.1 &   0.0 &   6.0 & PLUTO~\cite{pluto}   \\
31.100 &  3.660 & 362.0 &   5.1 &   0.0 &   3.5 & TASSO~\cite{tassoa}  \\
\hline
31.290 &  3.830 & 374.3 &   7.4 &   0.0 &   2.4 & JADE~\cite{jade}     \\
33.200 &  4.090 & 355.5 &   4.5 &   0.0 &   3.5 & TASSO~\cite{tassoa}  \\
33.790 &  3.860 & 324.1 &   1.8 &   3.0 &   1.6 & MARKJ~\cite{markj}   \\
33.800 &  3.740 & 313.8 &   2.7 &   1.9 &   1.7 & CELLO~\cite{cello}   \\
33.890 &  4.160 & 347.2 &   2.3 &   0.0 &   2.4 & JADE~\cite{jade}     \\
34.000 &  4.120 & 341.7 &   2.6 &   0.0 &   3.5 & TASSO~\cite{tassoa}  \\
34.500 &  3.930 & 316.6 &   5.1 &   0.0 &   2.4 & JADE~\cite{jade}     \\
34.610 &  3.780 & 302.6 &   0.8 &   3.0 &   1.6 & MARKJ~\cite{markj}   \\
34.700 &  4.080 & 325.0 &   2.2 &   0.0 &   3.5 & TASSO~\cite{tassoa}  \\
35.000 &  4.150 & 325.0 &   0.5 &   0.0 &   3.5 & TASSO~\cite{tassob}  \\
35.010 &  3.930 & 307.6 &   2.5 &   0.0 &   2.4 & JADE~\cite{jade}     \\
35.100 &  3.940 & 306.8 &   1.5 &   3.0 &   1.6 & MARKJ~\cite{markj}   \\
\hline
35.450 &  3.930 & 300.1 &   4.6 &   0.0 &   2.4 & JADE~\cite{jade}     \\
36.100 &  3.930 & 289.5 &   4.8 &   0.0 &   3.5 & TASSO~\cite{tassoa}  \\
36.310 &  3.880 & 282.5 &   4.2 &   3.0 &   1.6 & MARKJ~\cite{markj}   \\
36.380 &  3.710 & 269.1 &   5.8 &   0.0 &   2.4 & JADE~\cite{jade}     \\
\hline
37.400 &  3.590 & 246.6 &   9.3 &   3.0 &   1.6 & MARKJ~\cite{markj}   \\
38.300 &  3.890 & 254.9 &   2.6 &   1.7 &   1.7 & CELLO~\cite{cello}   \\
38.380 &  4.030 & 263.0 &   4.7 &   3.0 &   1.6 & MARKJ~\cite{markj}   \\
\hline
40.320 &  4.050 & 239.7 &   4.7 &   0.0 &   2.6 & JADE~\cite{jade}     \\
40.340 &  3.870 & 228.9 &   4.2 &   3.0 &   1.6 & MARKJ~\cite{markj}   \\
41.180 &  4.210 & 239.0 &   5.1 &   0.0 &   2.6 & JADE~\cite{jade}     \\
41.500 &  4.030 & 225.3 &   4.2 &   1.8 &   1.7 & CELLO~\cite{cello}   \\
41.500 &  4.440 & 248.3 &   4.5 &   3.0 &   1.6 & MARKJ~\cite{markj}   \\
42.500 &  3.890 & 207.5 &   5.2 &   3.0 &   1.6 & MARKJ~\cite{markj}   \\
42.550 &  4.200 & 223.5 &   5.1 &   0.0 &   2.6 & JADE~\cite{jade}     \\
\hline
43.460 &  3.750 & 191.4 &   4.7 &   3.0 &   1.6 & MARKJ~\cite{markj}   \\
43.500 &  3.970 & 202.2 &   2.0 &   1.4 &   1.7 & CELLO~\cite{cello}   \\
43.530 &  4.000 & 203.5 &   5.0 &   0.0 &   2.6 & JADE~\cite{jade}     \\
43.700 &  4.110 & 207.5 &   1.2 &   0.0 &   3.5 & TASSO~\cite{tassob}  \\
44.200 &  4.010 & 197.9 &   2.5 &   1.2 &   1.7 & CELLO~\cite{cello}   \\
44.230 &  4.150 & 204.6 &   1.9 &   3.0 &   1.6 & MARKJ~\cite{markj}   \\
44.410 &  3.980 & 194.6 &   5.1 &   0.0 &   2.6 & JADE~\cite{jade}     \\
\hline
45.480 &  4.170 & 194.5 &   4.5 &   3.0 &   1.6 & MARKJ~\cite{markj}   \\
45.590 &  4.400 & 204.3 &   4.8 &   0.0 &   2.6 & JADE~\cite{jade}     \\
46.000 &  4.090 & 186.6 &   5.2 &   1.9 &   1.7 & CELLO~\cite{cello}   \\
46.470 &  4.420 & 197.6 &   3.7 &   3.0 &   1.6 & MARKJ~\cite{markj}   \\
46.470 &  4.040 & 180.6 &   6.0 &   0.0 &   2.6 & JADE~\cite{jade}     \\
46.600 &  4.200 & 186.7 &   8.5 &   1.7 &   1.7 & CELLO~\cite{cello}   \\

\hline\hline
\end{tabular}
}
\end{center}
\end{table}

\begin{table}[htbp]
\begin{center}
\caption{\footnotesize The ratio $R$ and the effective Born hadronic cross 
section, $\sigma^0_{\rm had}$, from the TRISTAN experiments,
with increasing centre-of-mass energy ($\sqrt{s}$). The expected statistical 
($\sigma_{\rm stat}$), point-to-point systematic ($\sigma_{\rm ptp}$) and 
normalization systematic ($\Delta_{\rm norm}$) uncertainties are 
also given (in \%). The latter is correlated between all energy points 
in a given publication. An additional normalization error $\Delta_{\rm QED} 
= 0.1\%$, fully correlated between all measurements, is to be added to account 
for missing QED higher orders. The last column points to the original 
publication.
\label{tab:tristan}}
\vspace{5mm}
{\scriptsize
\begin{tabular}{|c|c|c|ccc|l|} \hline\hline
\rule{0pt}{5.0mm}
$\sqrt{s}$ (GeV) & Ratio $R$ & $\sigma^0_{\rm had}$ (pb) & 
$\sigma_{\rm stat}$ (\%) & 
$\sigma_{\rm ptp}$ (\%) & 
$\Delta_{\rm norm}$ (\%) & 
Reference \\ 
\hline\hline
50.000 &  4.530 & 175.2 &  12.7 &   2.3 &   2.7 & TOPAZ~\cite{topaza}  \\
50.000 &  4.400 & 170.2 &  11.2 &   4.0 &   0.7 & VENUS~\cite{venusa}  \\
50.000 &  4.500 & 174.1 &  10.5 &   2.8 &   1.6 & AMY~\cite{amy}       \\
52.000 &  4.530 & 162.1 &   4.6 &   1.1 &   2.7 & TOPAZ~\cite{topaza}  \\
52.000 &  4.700 & 168.2 &   6.2 &   4.0 &   0.7 & VENUS~\cite{venusa}  \\
52.000 &  4.289 & 153.5 &   4.7 &   2.2 &   1.6 & AMY~\cite{amy}       \\
54.000 &  4.979 & 165.4 &  10.9 &   3.4 &   2.7 & TOPAZ~\cite{topaza}  \\
54.000 &  4.688 & 155.7 &   9.2 &   1.8 &   1.6 & VENUS~\cite{venusb}  \\
54.000 &  4.725 & 156.9 &  12.8 &   3.4 &   1.6 & AMY~\cite{amy}       \\
55.000 &  4.639 & 148.6 &   5.4 &   1.4 &   2.7 & TOPAZ~\cite{topaza}  \\
55.000 &  4.317 & 138.3 &   7.2 &   1.8 &   1.6 & VENUS~\cite{venusb}  \\
55.000 &  4.632 & 148.4 &   5.2 &   1.4 &   1.6 & AMY~\cite{amy}       \\
\hline  
56.000 &  5.068 & 156.7 &   4.2 &   0.8 &   2.7 & TOPAZ~\cite{topaza}  \\
56.000 &  4.655 & 143.9 &   3.9 &   1.8 &   1.6 & VENUS~\cite{venusb}  \\
56.000 &  5.207 & 161.0 &   3.5 &   1.1 &   1.6 & AMY~\cite{amy}       \\
56.500 &  5.108 & 155.2 &   9.1 &   2.1 &   2.7 & TOPAZ~\cite{topaza}  \\
56.500 &  3.935 & 119.5 &  11.5 &   1.8 &   1.6 & VENUS~\cite{venusb}  \\
56.500 &  5.324 & 161.7 &   8.7 &   2.5 &   1.6 & AMY~\cite{amy}       \\
57.000 &  5.147 & 153.7 &   4.7 &   1.1 &   2.7 & TOPAZ~\cite{topaza}  \\
57.000 &  4.983 & 148.7 &   4.3 &   1.8 &   1.6 & VENUS~\cite{venusb}  \\
57.000 &  4.903 & 146.4 &   4.5 &   1.3 &   1.6 & AMY~\cite{amy}       \\
\hline  
57.370 &  4.432 & 130.6 &  10.4 &   0.0 &   2.2 & TOPAZ~\cite{topazb}  \\
57.770 &  4.878 & 141.8 &   0.9 &   0.9 &   0.0 & VENUS~\cite{venusc}  \\
57.770 &  4.940 & 143.6 &   1.0 &   0.0 &   2.2 & TOPAZ~\cite{topazc}  \\
57.970 &  4.832 & 139.5 &   9.4 &   0.0 &   2.2 & TOPAZ~\cite{topazb}  \\
58.220 &  4.727 & 135.3 &   9.1 &   0.0 &   2.2 & TOPAZ~\cite{topazb}  \\
58.290 &  5.336 & 152.4 &   8.0 &   1.7 &   2.7 & TOPAZ~\cite{topaza}  \\
58.470 &  4.291 & 121.8 &  10.7 &   0.0 &   2.2 & TOPAZ~\cite{topazb}  \\
58.500 &  4.909 & 139.2 &   8.9 &   1.8 &   1.6 & VENUS~\cite{venusb}  \\
58.500 &  5.303 & 150.4 &  10.4 &   2.0 &   1.6 & AMY~\cite{amy}       \\
58.720 &  4.811 & 135.4 &   8.3 &   0.0 &   2.2 & TOPAZ~\cite{topazb}  \\
58.970 &  5.582 & 155.8 &   8.0 &   0.0 &   2.2 & TOPAZ~\cite{topazb}  \\
59.000 &  4.848 & 135.2 &   9.7 &   1.8 &   1.6 & VENUS~\cite{venusb}  \\
59.000 &  5.409 & 150.8 &  10.9 &   2.8 &   1.6 & AMY~\cite{amy}       \\
59.050 &  6.055 & 168.5 &   9.8 &   1.8 &   1.6 & VENUS~\cite{venusb}  \\
59.050 &  6.582 & 183.2 &  10.8 &   2.6 &   1.6 & AMY~\cite{amy}       \\
59.060 &  5.735 & 159.6 &   7.1 &   2.1 &   2.7 & TOPAZ~\cite{topaza}  \\
\hline  
59.220 &  5.084 & 140.7 &   9.3 &   0.0 &   2.2 & TOPAZ~\cite{topazb}  \\
59.470 &  5.447 & 149.5 &   9.8 &   0.0 &   2.2 & TOPAZ~\cite{topazb}  \\
59.840 &  4.717 & 127.9 &   8.1 &   0.0 &   2.2 & TOPAZ~\cite{topazb}  \\
60.000 &  5.305 & 143.1 &   5.4 &   1.3 &   2.7 & TOPAZ~\cite{topaza}  \\
60.000 &  5.274 & 142.2 &   4.7 &   1.8 &   1.6 & VENUS~\cite{venusb}  \\
60.000 &  5.809 & 156.7 &   4.7 &   1.3 &   1.6 & AMY~\cite{amy}       \\
60.800 &  5.653 & 148.5 &   4.8 &   1.1 &   2.7 & TOPAZ~\cite{topaza}  \\
60.800 &  5.680 & 149.2 &   4.1 &   1.8 &   1.6 & VENUS~\cite{venusb}  \\
60.800 &  5.544 & 145.7 &   5.2 &   1.9 &   1.6 & AMY~\cite{amy}       \\
\hline  
61.400 &  5.852 & 150.8 &   5.1 &   1.4 &   2.7 & TOPAZ~\cite{topaza}  \\
61.400 &  4.990 & 128.6 &   4.4 &   1.8 &   1.6 & VENUS~\cite{venusb}  \\
61.400 &  5.410 & 139.4 &   5.0 &   1.4 &   1.6 & AMY~\cite{amy}       \\
63.600 &  6.126 & 147.2 &  10.7 &   1.8 &   1.6 & VENUS~\cite{venusb}  \\
\hline\hline
\end{tabular}
}
\end{center}
\end{table}

The $R$ and $\sigma^0_{\rm had}$ data are listed in Table~\ref{tab:petra}
(PEP, PETRA) and in Table~\ref{tab:tristan} (TRISTAN), as obtained 
from a comparison of two recent compilations~\cite{pdgfiles,whalley} and 
%the original publications~{\catcode`@=11 [\hbox{\b@markii}--\hbox{\b@venusc}]}.
the original publications~\cite{markii}--\cite{venusc}.
In these tables, only the final -- and most accurate -- result for each 
experiment and each centre-of-mass energy is reported. (Superseded data 
are reported in both Refs.~\cite{pdgfiles} and~\cite{whalley}, but are not 
always clearly flagged as such therein.)

Other refinements were considered in this letter for a rigorous statistical
treatment of the data, and are described in the following. First, in each 
experiment, the systematic uncertainty was divided into a point-to-point 
contribution, $\sigma_{\rm ptp}$, and an overall normalization error, 
$\Delta_{\rm norm}$, as is done in most of the orginal publications. 
The point-to-point systematic uncertainties are uncorrelated (related to, 
{\it e.g.,} the limited simulated statistics, or the statistical uncertainty 
on the measured luminosity), are assumed to have a Gaussian probability 
density function and are taken directly from the original publications. 

In contrast, the overall normalization error definition varies among the 
publications, being either the largest possible variation interval 
({\it e.g.,} between several sets of selection criteria, different ways 
of determining the luminosity, or various quark fragmentation models) 
or half this interval. Here, the definition was unified in such a way 
that the overall normalization can vary by $-\Delta_{\rm norm}$ and 
$+\Delta_{\rm norm}$, with a uniform probability over the whole interval. 
This overall normalization error is 100\% correlated between the different 
centre-of-mass energy points reported in each given publication.

Third, the published values of $\Delta_{\rm norm}$ often contain an 
estimate of the effect of missing higher-order QED corrections in the ISR
unfolding procedure, at the level of a couple of percent. Indeed, at the 
time of PEP, PETRA and TRISTAN, the Monte Carlo programs used to simulate 
the $\epemto\ \qqbar$ and $\epemto\ \epem$ processes were limited 
to ${\cal O}(\alpha_{\rm QED})$. The missing orders have a potential 
effect on the measured value of $\sigma^0_{\rm had}$ via the prediction 
of both the hadronic cross section and the Bhabha scattering cross section: 
the former is used to correct the measured $\sigma_{\rm had}$ for QED 
effects, and the latter to determine the integrated luminosity. Altogether, 
the published cross section values would have to be corrected as follows,
\begin{equation}
\sigma^0_{\rm had} \longrightarrow \sigma^0_{\rm had} \times
{\sigma_{\rm ee}^{\rm (all)} \over \sigma_{\rm ee}^{(1)}}
{\sigma_{\rm had}^{\rm (1)} \over \sigma_{\rm had}^{\rm (all)}},
\end{equation}
where the indices (1) and (all) refer to the cross section prediction 
up to the QED first order (used in the original publications) and with 
all orders, respectively. 

With the programs that have been developed for LEP, it is now possible to 
evaluate this correction with a better accuracy than that assumed twenty years 
ago. The $\epemto\ \qqbar$ and $\epemto\ \epem$ cross sections were determined 
here with and without QED higher orders by {\tt ZFITTER}~\cite{zfitter} with 
an emulation of the kinematical cuts described in the original publications. 
It was found that the corrections to Bhabha scattering and hadron production 
essentially cancel in the ratio of Eq.~2. The remaining contribution of 
QED higher orders is at the 0.1\% level, almost independently of the 
event selection and the centre-of-mass energy. 

The large uncertainties related to the missing QED higher orders were 
therefore taken out from the original values of $\Delta_{\rm norm}$. 
While the aforementioned 0.1\% contribution could be simply corrected for
in $\sigma_{\rm had}^0$, 
a new normalization error $\Delta_{\rm QED} = 0.1\%$ was added instead 
(assumed to be 100\% correlated between all PEP, PETRA and TRISTAN 
measurements) to conservatively account for the yet missing orders 
in {\tt ZFITTER}. 

Finally, early TRISTAN data~\cite{amy,topaza,venusa,venusb}
are also corrected in the original publications for other electroweak 
effects, dominated by the top quark contribution (with a 
$(m_{\rm top}/m_{\rm Z})^2$ dependence at first order). These small 
corrections (between $+0.1\%$ and $+0.7\%$ at $\sqrt{s}=60$\,GeV, 
depending on the top quark mass chosen to determine the correction)
were unfolded here {\it (i)} to have a consistent data set to work with; 
and {\it (ii)} for a sound comparison with the {\tt ZFITTER} prediction, 
which includes first- and higher-order electroweak contributions as well. 
The latest 
TRISTAN data~\cite{topazb,topazc,venusc} were, more adequately, 
corrected for QED effects only. The electroweak  effect correction needs 
therefore not be unfolded in that case.

For practical reasons, the measurements of Tables~\ref{tab:petra} 
and~\ref{tab:tristan} were clustered in few centre-of-mass bins 
as indicated by the horizontal separation lines in these two tables. 
The ratio $R$ values were averaged in each bin according to the total  
uncertainties, {\it i.e.}, with a weight proportional to the inverse 
of $\sigma_{\rm tot}^2 = R^2 \times \left( 
\sigma_{\rm stat}^2 + \sigma^2_{\rm ptp} + \Delta^2_{\rm norm}/3 + 
\Delta^2_{\rm QED}/3 \right)$. The corresponding averaged Born effective
cross sections ($\sigma^0_{\rm had}$) and centre-of-mass energy values 
are displayed in Table~\ref{tab:average}. The $R$ values found for PEP/PETRA 
were found to agree with those of an earlier combination~\cite{cello}.
The effective Born hadronic cross section ($\sigma^0_{\rm th}$) 
predicted by {\tt ZFITTER}~\cite{zfitter} is also 
shown in Table~\ref{tab:average}.

\begin{table}[h]
\begin{center}
\caption{\footnotesize The ratio $R$ and the effective Born hadronic cross 
section, $\sigma^0_{\rm had}$, from PEP, PETRA, TRISTAN, as 
a function of the centre-of-mass energy ($\sqrt{s}$), averaged in 
$\sim 2$-GeV-wide centre-of-mass-energy bins. The hadronic cross-section 
prediction, $\sigma^0_{\rm th}$, is also shown. The last column displays 
the number 
of measurements used for each entry.
\label{tab:average}}
\vspace{5.5mm}
\begin{tabular}{|c|c|c|c|c|} \hline\hline
\rule{0pt}{4.0mm}
$\sqrt{s}$ (GeV) & Ratio $R$ & $\sigma^0_{\rm had}$ (pb) & $\sigma^0_{\rm th}$ (pb) & $N_{\rm pts}$ \\ 
\hline\hline
\rule{0pt}{4.2mm}
  21.995 & $  3.843 \pm   0.067$ & $ 754.1 \pm   13.1$ & 763.1&   5  \\
\rule{0pt}{4.2mm}
  25.003 & $  4.047 \pm   0.167$ & $ 616.4 \pm   25.4$ & 592.2&   3  \\
\rule{0pt}{4.2mm}
  28.932 & $  3.945 \pm   0.045$ & $ 450.3 \pm    5.2$ & 444.4&   5  \\
\rule{0pt}{4.2mm}
  30.570 & $  3.929 \pm   0.086$ & $ 402.2 \pm    8.8$ & 399.1&   6  \\
\rule{0pt}{4.2mm}
  34.408 & $  3.996 \pm   0.038$ & $ 323.9 \pm    3.1$ & 317.6&  12  \\
\rule{0pt}{4.2mm}
  36.022 & $  3.871 \pm   0.102$ & $ 286.6 \pm    7.5$ & 291.0&   4  \\
\rule{0pt}{4.2mm}
  38.237 & $  3.894 \pm   0.105$ & $ 255.9 \pm    6.9$ & 260.2&   3  \\
\rule{0pt}{4.2mm}
  41.329 & $  4.083 \pm   0.081$ & $ 230.3 \pm    4.6$ & 225.7&   7  \\
\rule{0pt}{4.2mm}
  43.825 & $  4.027 \pm   0.051$ & $ 202.2 \pm    2.6$ & 203.7&   7  \\
\rule{0pt}{4.2mm}
  46.038 & $  4.234 \pm   0.098$ & $ 192.9 \pm    4.5$ & 187.6&   6  \\
\rule{0pt}{4.2mm}
  53.097 & $  4.527 \pm   0.097$ & $ 155.8 \pm    3.3$ & 153.6&  12  \\
\rule{0pt}{4.2mm}
  56.432 & $  4.964 \pm   0.087$ & $ 151.1 \pm    2.6$ & 145.4&   9  \\
\rule{0pt}{4.2mm}
  57.867 & $  4.926 \pm   0.046$ & $ 142.7 \pm    1.3$ & 143.4&  16  \\
\rule{0pt}{4.2mm}
  60.264 & $  5.456 \pm   0.107$ & $ 145.8 \pm    2.9$ & 142.2&   9  \\
\rule{0pt}{4.2mm}
  61.521 & $  5.378 \pm   0.156$ & $ 138.0 \pm    4.0$ & 142.8&   4  \\
\hline\hline
\end{tabular}
\end{center}
\end{table}

\noindent
Computer-readable files for these data will be transmitted
to the Review of Particle Physics and are available at
{\tt http://janot.web.cern.ch/janot/HadronicData/}.

\begin{figure}[htbp]
\begin{picture}(160,120)
\put(15,-2){\epsfxsize130mm\epsfbox{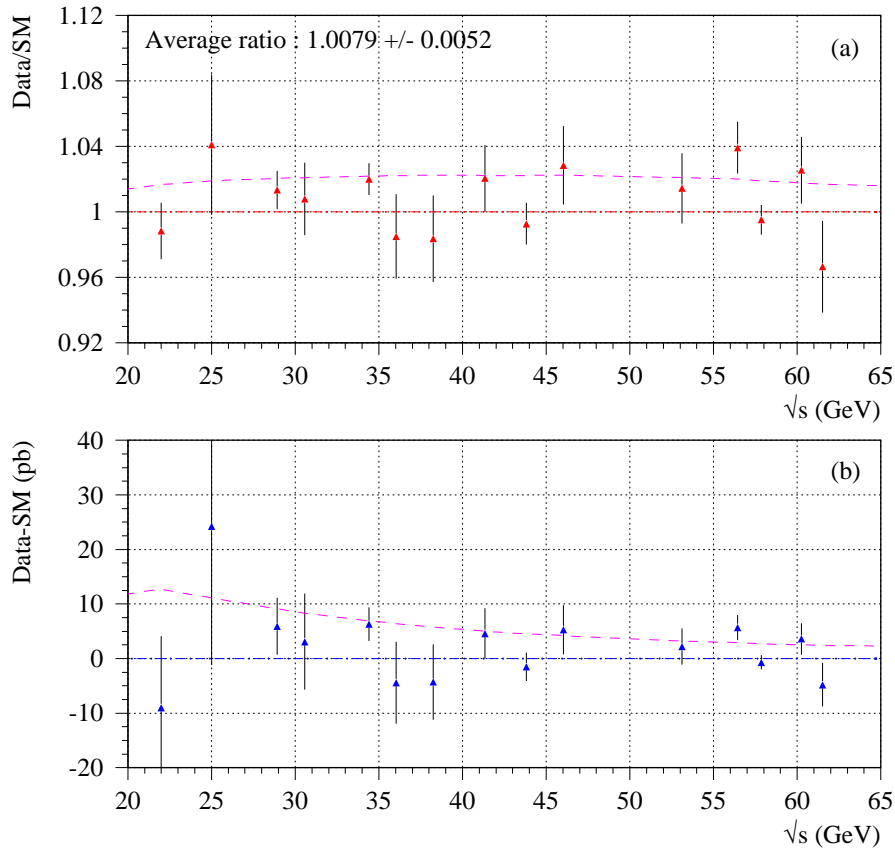}}
\end{picture}
\caption[ ]
{\protect\footnotesize Ratio (a) and difference (b) of the effective 
Born hadronic cross 
section measurements and the {\tt ZFITTER} prediction as a function 
of the centre-of-mass energy, for PEP, PETRA and TRISTAN data, 
rebinned as explained in the text. The dash-dotted line indicates
the standard model prediction, and the dashed curve the additional contribution
of sbottom pair production with $\msbot\ = 6\,\gevcc$ and with a vanishing 
coupling to the Z. 
\label{fig:ratio}}
\end{figure}

The ratio and the difference of these measured cross sections 
and those predicted by {\tt ZFITTER} are displayed in Fig.~\ref{fig:ratio} 
as a function of the centre-of-mass energy. When no systematic uncertainties 
are assigned to the theoretical prediction, the average ratio appears to 
exceed the prediction by $(0.79 \pm 0.52)\%$, {\it i.e.,} by 1.5 standard
deviations. This excess is, however, about 2.4 standard deviations
below the prediction of an additional light sbottom pair production
(here with a mass of 6\,\gevcc), 
which would amount to about 2\% of the total cross section.

The experimental correlations between the different bins, essential 
for a rigourous statistical treatment of the data, were determined 
following the lines of Ref.~\cite{cello}. In practice, a Monte Carlo 
technique relying on the generation of many gedanken experiments 
was used to determine the probability density functions of the measured
$R$ ratio values listed in Tables~\ref{tab:petra} and~\ref{tab:tristan}.
In each gedanken experiment, 108 $R$ values were generated around 
the measured central value, smeared by {\it (i)} a Gaussian distribution 
with a width equal to the quadratic sum of $\sigma_{\rm stat}$ and 
$\sigma_{\rm ptp}$; {\it (ii)} a uniform distribution in the
$\left[-\Delta_{\rm norm}, +\Delta_{\rm norm}\right]$ interval, 
identical for all energy points of a given publication; and 
{\it (iii)} a uniform distribution in the
$\left[-\Delta_{\rm QED}, +\Delta_{\rm QED}\right]$ interval, 
identical for all 108 measurements. 

\eject
As above, an average value $R_i$ was determined in each centre-of-mass-energy 
bin $i$ for each gedanken experiment. This allowed the $R_i$ values of 
Table~\ref{tab:average} and their uncertainties to be confirmed, when 
averaging over a large number of gedanken experiments. Similarly, the 
uncertainties of the cross-products $R_i \times R_j$ led to the correlation 
matrices shown in Tables~\ref{tab:petracor} and~\ref{tab:tristancor}, 
for PEP and PETRA on the one hand, and for TRISTAN on the other. The
cross-correlations between PEP, PETRA and TRISTAN (induced solely by 
$\Delta_{\rm QED}$) were found to be smaller than $5\,10^{-4}$ and 
were therefore neglected in the following.

\begin{table}[htbp]
\begin{center}
\caption{\footnotesize The correlations between the ten PEP and PETRA 
centre-of-mass energy bins ($\sqrt{s}$ in GeV).
\label{tab:petracor}}
\vspace{5mm}
{\footnotesize
\begin{tabular}{|c|cccccccccc|} \hline\hline

$\sqrt{s}$ %(GeV) 
&21.994 &25.003 &28.932 &30.572 &34.409 &36.027 &38.231 &41.325 &43.824 &46.042
\\
\hline 21.994&
1.000 & 0.034 & 0.003 & 0.043 & 0.237 & 0.060 & 0.028 & 0.084 & 0.227 & 0.066 
\\ 25.003&
0.034 & 1.000 & 0.004 & 0.053 & 0.096 & 0.055 & 0.012 & 0.053 & 0.029 & 0.041 
\\ 28.932&
0.003 & 0.004 & 1.000 & 0.032 & 0.017 & 0.009 & 0.000 & 0.003 & 0.001 & 0.002 
\\ 30.572&
0.043 & 0.053 & 0.032 & 1.000 & 0.198 & 0.105 & 0.009 & 0.057 & 0.030 & 0.043 
\\ 34.409&
0.237 & 0.096 & 0.017 & 0.198 & 1.000 & 0.185 & 0.032 & 0.148 & 0.258 & 0.113 
\\ 36.027&
0.060 & 0.055 & 0.009 & 0.105 & 0.185 & 1.000 & 0.010 & 0.096 & 0.048 & 0.071 
\\ 38.231&
0.028 & 0.012 & 0.000 & 0.009 & 0.032 & 0.010 & 1.000 & 0.030 & 0.054 & 0.030 
\\ 41.325&
0.084 & 0.053 & 0.003 & 0.057 & 0.148 & 0.096 & 0.030 & 1.000 & 0.078 & 0.098 
\\ 43.824&
0.227 & 0.029 & 0.001 & 0.030 & 0.258 & 0.048 & 0.054 & 0.078 & 1.000 & 0.066 
\\ 46.042&
0.066 & 0.041 & 0.002 & 0.043 & 0.113 & 0.071 & 0.030 & 0.098 & 0.066 & 1.000 
\\
\hline\hline
\end{tabular}
}
\end{center}
\end{table}

\begin{table}[htbp]
\begin{center}
\caption{\footnotesize The correlations between the five TRISTAN 
centre-of-mass energy bins ($\sqrt{s}$ in GeV).
\label{tab:tristancor}}
\vspace{5mm}
{\footnotesize
\begin{tabular}{|c|ccccc|} \hline\hline
$\sqrt{s}$ & 
53.141 & 56.436 & 57.863 & 60.253 & 61.519 \\
\hline 
53.141 & 1.000 & 0.101 & 0.014 & 0.077 & 0.057 \\
56.436 & 0.101 & 1.000 & 0.017 & 0.093 & 0.073 \\
57.863 & 0.014 & 0.017 & 1.000 & 0.022 & 0.010 \\
60.253 & 0.077 & 0.093 & 0.022 & 1.000 & 0.058 \\
61.519 & 0.057 & 0.073 & 0.010 & 0.058 & 1.000 \\
\hline\hline
\end{tabular}
}
\end{center}
\end{table}

\subsection{The LEP\,1 and SLC data} 
\label{sec:lepslc}

\begin{table}[htbp]
\begin{center}
\caption{\footnotesize Precise LEP and SLC measurements of the Z lineshape 
parameters ($\Gamma_{\rm Z}$, $R_\ell$, $\sigma_{\rm had}$), of $g_V/g_A$ 
and of $R_{\rm b}$, together with their correlation matrix. The last 
two measurements have been taken here as uncorrelated with the first 
three~\cite{bolek}. The standard model prediction formula are given in 
Ref.~\cite{janot}.
\label{tab:lepslc}}
\vspace{3mm}
{\footnotesize
\begin{tabular}{|l|l||rrrrr|} \hline\hline
Observable & Measurement & \multicolumn{5}{|c|}{Correlation matrix} \\ 
\hline
$\Gamma_{\rm Z}$ & $2495.2 \pm\ 2.3$\,MeV  & $1.000$ & & & &     \\ \hline
$R_\ell$         & $20.767 \pm\ 0.025$     & $+0.004$ & $1.000$ & & & 
\\ \hline
$\sigma_{\rm had}$ &$41.541 \pm\ 0.037$\,nb & $-0.297$ & $+0.183$ & $1.000$ & &
\\ \hline
$g_V/g_A$ &  $0.07408 \pm\ 0.00068$  & $0.000$ & $0.000$ & $0.000$ & $1.000$ &
\\ \hline
$R_{\rm b}$ &$0.21638 \pm\ 0.00066$ & $0.000$ & $0.000$ & $0.000$ & $0.000$ & 
$1.000$     \\ \hline\hline
\end{tabular}
}
\end{center}
\end{table}
\noindent

The precise measurements of LEP and SLC and their correlations~\cite{lep} 
are summarized in Table~\ref{tab:lepslc}. 
Most of these Z observables would be modified in case 
of an additional New Physics contribution to hadronic Z decays. Let \eNPh\ 
be the ratio of this new partial width $\Gamma_{\rm NP}$ to the total 
decay width of the Z without this new contribution. As was shown 
in Ref.~\cite{janot}, the Z total width $\Gamma_{\rm Z}$, the ratio 
$R_\ell$ of the hadronic to the leptonic branching fractions, 
and the peak cross section $\sigma^0_{\rm had}$ are modified as follows,

\begin{eqnarray}
\Gamma_{\rm Z} & \longrightarrow & \Gamma_{\rm Z}\,\left(1+1.00\eNPh\right),
\ \ \ \ \ \left[ \Gamma_{\rm Z} + \Gamma_{\rm NP} \right] 
\\ 
\nonumber \\
R_\ell & \longrightarrow & R_\ell\,\left(1+1.43\eNPh\right),
\ \ \ \ \ \left[ \left(\Gamma_{\rm had} + \Gamma_{\rm NP}\right)
/\Gamma_\ell \right] \\ 
\nonumber \\
\sigma^0_{\rm had} & \longrightarrow & \sigma^0_{\rm had}\,
\left(1 - 0.57\eNPh\right).
\ \ \ \left[ \frac{12\pi}{m_{\rm Z}^2}
 \frac{\Gamma_{\rm ee}(\Gamma_{\rm had} + \Gamma_{\rm NP})}
 {(\Gamma_{\rm Z} + \Gamma_{\rm NP})^2} \right]
\end{eqnarray}

In Ref.~\cite{janot}, the new hadronic decay channel considered was 
flavour-democratic. The individual branching fractions into the different 
quark flavours were therefore not modified by this new contribution. 
In the case of a sbottom pair production with hadronic R-parity-violating 
decays into light quarks exclusively, the ratio of the ${\rm b\bar b}$ 
branching ratio to the hadronic branching ratio, $R_{\rm b}$, is also 
modified according to

\begin{equation}
R_{\rm b}  \longrightarrow  R_{\rm b}\,\left(1-1.43\eNPh\right),
\ \ \ \ \ \left[ \Gamma_{\rm b\bar b} / \left(\Gamma_{\rm had} + 
\Gamma_{\rm NP}\right) \right],
\end{equation}
while $(g_V/g_A)$ remains untouched.

These observables would also be modified by the virtual 
corrections arising from the New Physics responsible for the additional 
hadronic contribution. As in Ref.~\cite{janot}, the value of \eNPh\ 
was fitted to the measurement of the five observables together with 
the generic contribution of these virtual effects. The result is

\begin{equation}
\eNPh\ = (-0.56 \pm 0.80)\times 10^{-3},
\end{equation}
which corresponds to an additional hadronic contribution of 

\begin{equation}
\sigma^{\rm NP}_{\rm had}(\mZ) = -24 \pm 36\,{\rm pb}.
\end{equation}

\begin{figure}[htbp]
\begin{picture}(160,56)
\put(27,-6){\epsfxsize105mm\epsfbox{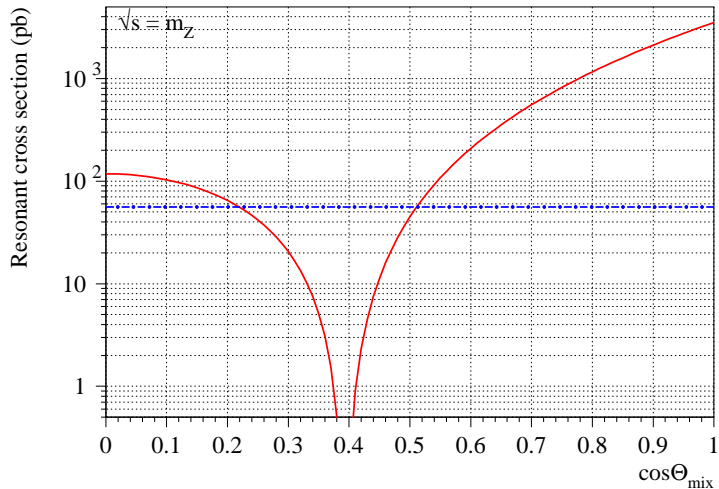}}
\end{picture}
\caption[ ]
{\protect\footnotesize The resonant contribution of the sbottom pair 
production cross section with $\msbot\ = 6\,\gevcc$, at $\sqrt{s}=\mZ$,
as a function of  $\cos\theta_{\rm mix}$ (full curve). The dash-dotted line 
indicates the 95\%\,C.L. upper limit on this cross section when the 
sbottom decays into light quarks exclusively. 
\label{fig:kraml}}
\end{figure}

It allows a 95\%\,C.L. upper limit of 56\,pb to be set on the cross section, 
at the Z peak, of any additional hadronic contribution to the Z decays into 
light quarks only. The resonant contribution of the sbottom pair production 
cross section~\cite{kraml} with $\msbot\ = 6\,\gevcc$ is shown in 
Fig.~\ref{fig:kraml} as a function of the mixing angle $\cos\theta_{\rm mix}$ 
between the two sbottom states $\tilde b_L$ and $\tilde b_R$, superpartners
of the left-handed and right-handed bottom quarks, respectively. For 
$\cos\theta_{\rm mix} \simeq 0.39$, the coupling between the Z and the 
lighter sbottom vanishes. 

For $\msbot\ = 6\,\gevcc$, the Z data allow all 
values of $\cos\theta_{\rm mix}$ below 0.22 and above 0.52 to be excluded at 
the 95\% confidence level. These data are therefore incompatible with a light 
sbottom pair production, unless the coupling to the Z is negligibly small.

\eject\null

\subsection{The LEP\,2 data}
\label{sec:lep2}

\begin{figure}[htbp]
\begin{picture}(160,110)
\put(15,-4){\epsfxsize130mm\epsfbox{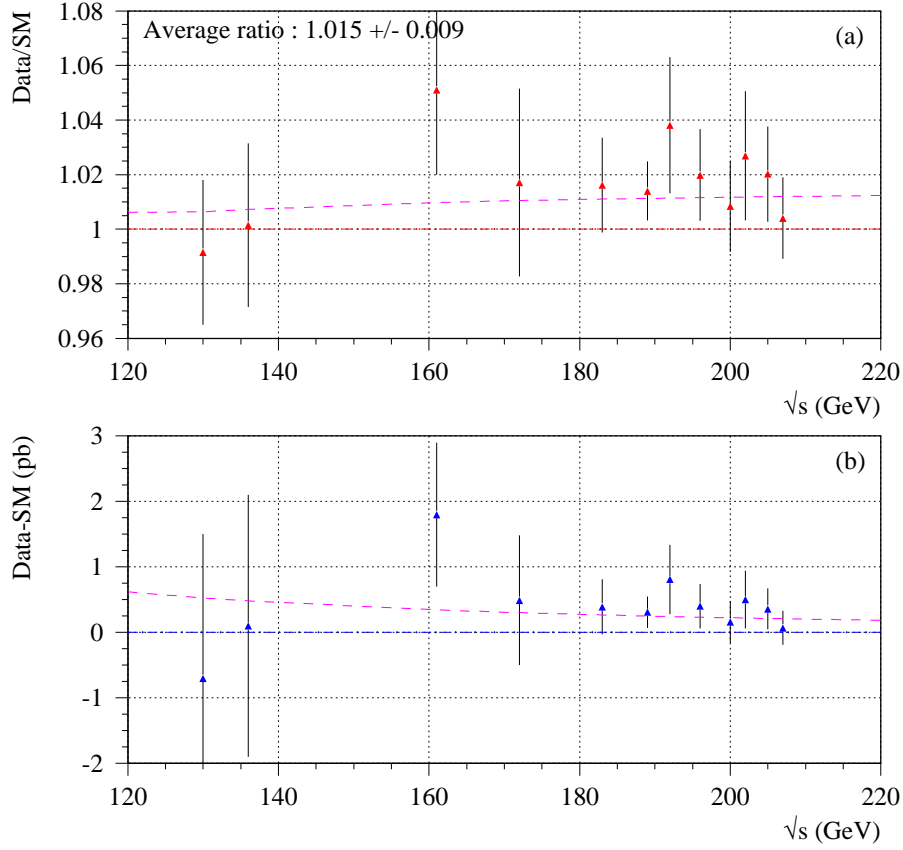}}
\end{picture}
\caption[ ]
{\protect\footnotesize Ratio (a) and difference (b) of the 
hadronic cross section measurements and the standard model prediction as 
a function of the centre-of-mass energy, for the LEP\,2 data.
The dash-dotted line indicates the standard model prediction, and the 
dashed curve the additional contribution of sbottom pair production 
with $\msbot\ = 6\,\gevcc$ and $\cos\theta_{\rm mix}=0.39$.
\label{fig:lep2data}}
\end{figure}

The preliminary LEP\,2 hadronic cross section data were taken 
from Ref.~\cite{lep}. The measured cross sections $\sigma_{\rm had}$ 
and the standard model predictions $\sigma_{\rm th}$ are 
summarized in Table~\ref{tab:lep2data}. These data are displayed in 
Fig.~\ref{fig:lep2data} and the correlation matrix is given in 
Table~\ref{tab:lep2cor}.
\begin{table}[h]
\begin{center}
\caption{\footnotesize The hadronic cross section, $\sigma_{\rm had}$, 
measured at the twelve LEP\,2 centre-of-mass energies, and the predictions
in the standard model, $\sigma_{\rm th}$. These data are still preliminary.
\label{tab:lep2data}}
\vspace{5mm}
{\small
\begin{tabular}{|c|c|c|} \hline\hline
\rule{0pt}{2mm}
$\sqrt{s}$ (GeV) & $\sigma_{\rm had}$ (pb) & $\sigma_{\rm th}$ (pb) \\
\hline
%\rule{0pt}{2mm}
130 & $82.1  \pm  2.2$ & 82.8  \\ 
%\rule{0pt}{2mm}
136 & $66.7  \pm  2.0$ & 66.6  \\
%\rule{0pt}{2mm}
161 & $37.0  \pm  1.1$ & 35.2  \\ 
%\rule{0pt}{2mm}
172 & $29.23 \pm 0.99$ & 28.74 \\ 
%\rule{0pt}{2mm}
183 & $24.59 \pm 0.42$ & 24.20 \\ 
%\rule{0pt}{2mm}
189 & $22.47 \pm 0.24$ & 22.16 \\
%\rule{0pt}{2mm}
192 & $22.05 \pm 0.53$ & 21.24 \\
%\rule{0pt}{2mm}
196 & $20.53 \pm 0.34$ & 20.13 \\
%\rule{0pt}{2mm}
200 & $19.25 \pm 0.32$ & 19.09 \\
%\rule{0pt}{2mm}
202 & $19.07 \pm 0.44$ & 18.57 \\
%\rule{0pt}{2mm}
205 & $18.17 \pm 0.31$ & 17.81 \\
%\rule{0pt}{2mm}
207 & $17.49 \pm 0.26$ & 17.42 \\
\hline\hline
\end{tabular}
}
\end{center}
\end{table}
When no systematic uncertainties are assigned to the theoretical prediction, 
the average ratio appears to exceed the prediction by $(1.5 \pm 0.9)\%$, 
{\it i.e.,} by 1.7 standard deviations. This excess, although not significant,
is compatible with, and actually slightly larger than an additional light 
sbottom pair production with $\cos\theta_{\rm mix}=0.39$. The latter 
would amount to about 1\% of the total cross section in this 
centre-of-mass energy range.

\begin{table}[htbp]
\begin{center}
\caption{\footnotesize The correlations between the twelve LEP\,2 
centre-of-mass energy bins ($\sqrt{s}$ in GeV).
\label{tab:lep2cor}}
\vspace{5mm}
{\footnotesize
\begin{tabular}{|c|cccccccccccc|} \hline\hline

$\sqrt{s}$ 
& 130 & 136 & 161 & 172 & 183 & 189 & 192 & 196 & 200 & 202 & 205 & 207 
\\
\hline 130 &
1.000 & 0.071 & 0.080 & 0.072 & 0.114 & 0.146 & 0.077 & 0.105 & 0.120 & 0.086 & 0.117 & 0.138 
\\ 136 &
     0.071 & 1.000 & 0.075 & 0.067 & 0.106 & 0.135 & 0.071 & 0.097 & 0.110 & 0.079 & 0.109 & 0.128 
\\ 161 &
     0.080 & 0.075 & 1.000 & 0.077 & 0.120 & 0.153 & 0.080 & 0.110 & 0.125 & 0.090 & 0.124 & 0.145 
\\ 172 &
     0.072 & 0.067 & 0.077 & 1.000 & 0.108 & 0.137 & 0.072 & 0.099 & 0.112 & 0.081 & 0.111 & 0.130 
\\ 183 &
     0.114 & 0.106 & 0.120 & 0.108 & 1.000 & 0.223 & 0.117 & 0.158 & 0.182 & 0.129 & 0.176 & 0.208 
\\ 189 &
     0.146 & 0.135 & 0.153 & 0.137 & 0.223 & 1.000 & 0.151 & 0.206 & 0.235 & 0.168 & 0.226 & 0.268 
\\ 192 &
     0.077 & 0.071 & 0.080 & 0.072 & 0.117 & 0.151 & 1.000 & 0.109 & 0.126 & 0.090 & 0.118 & 0.138 
\\ 196 &
     0.105 & 0.097 & 0.110 & 0.099 & 0.158 & 0.206 & 0.109 & 1.000 & 0.169 & 0.122 & 0.162 & 0.190 
\\ 200 &
     0.120 & 0.110 & 0.125 & 0.112 & 0.182 & 0.235 & 0.126 & 0.169 & 1.000 & 0.140 & 0.184 & 0.215 
\\ 202 &
     0.086 & 0.079 & 0.090 & 0.081 & 0.129 & 0.168 & 0.090 & 0.122 & 0.140 & 1.000 & 0.132 & 0.153 
\\ 205 &
     0.117 & 0.109 & 0.124 & 0.111 & 0.176 & 0.226 & 0.118 & 0.162 & 0.184 & 0.132 & 1.000 & 0.213 
\\ 207 &
     0.138 & 0.128 & 0.145 & 0.130 & 0.208 & 0.268 & 0.138 & 0.190 & 0.215 & 0.153 & 0.213 & 1.000
\\

\hline\hline
\end{tabular}
}
\end{center}
\end{table}

\section{Global fit}
\label{sec:fits}

When no systematic uncertainties are assigned to the standard model 
prediction, the data can be combined in a global negative log-likelihood 
${\cal L}(\cos\theta_{\rm mix},\alpha)$ as follows, 
\begin{equation}
{\cal L}(\cos\theta_{\rm mix},\alpha) = 
{1\over 2}\sum_{i,j=1}^N \Delta_i S_{ij}^{-1} \Delta_j 
{\rm \ \ with \ \ }
\Delta_i = \sigma_{{\rm had},i} - \left[ \sigma_{{\rm th},i} + \alpha 
\sigma_{{\rm NP},i}(\msbot,\cos\theta_{\rm mix}) \right],
\end{equation}
where $S$ is the covariance matrix of the $N$ ($=28$) measurements of 
PEP, PETRA, TRISTAN, LEP\,1, SLC and LEP\,2 as compiled 
in Section~\ref{sec:data}, $\theta_{\rm mix}$ is the mixing angle in the 
sbottom sector and $\alpha$ is an arbitrary normalization constant of 
the sbottom pair production cross section, $\sigma_{{\rm NP},i}$. 
The likelihood is then minimized with respect to $\cos\theta_{\rm mix}$ and to 
$\alpha$ to find the best fit to the data. A fitted value of $\alpha$
compatible with unity and incompatible with 0.~would be the sign of New 
Physics, while a value compatible with 0., but incompatible with 1., 
would allow this New Physics to be excluded with a certain level of 
confidence. (This same technique
can be applied for any kind of New Physics leading to hadronic final 
states in \epem\ collisions.) 

\begin{figure}[htbp]
\begin{picture}(160,100)
\put(22,-7){\epsfxsize115mm\epsfbox{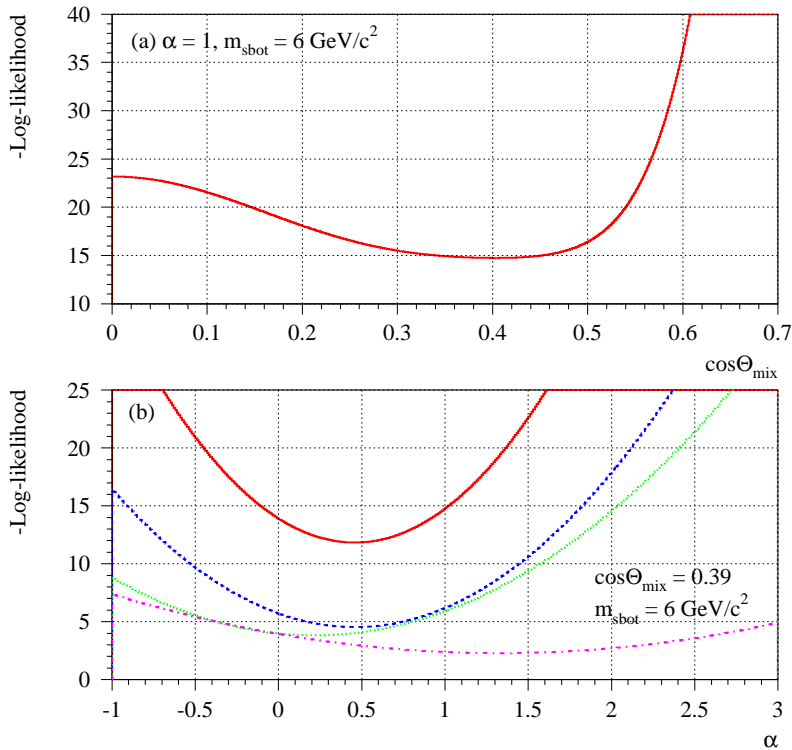}}
\end{picture}
\caption[ ]
{\protect\footnotesize The negative log-likelihood with $\msbot\ = 6\,\gevcc$
(a) as a function of $\cos\theta_{\rm mix}$ for $\alpha=1$; 
and (b) as a function of $\alpha$ with $\cos\theta_{\rm mix} = 0.39$ for 
the combined data (full curve), PEP/PETRA (dashed curve), 
TRISTAN (dotted curve) and LEP\,2 (dash-dotted curve).
\label{fig:mixing}}
\end{figure}

For $\alpha=1$ and $\msbot\ = 6\,\gevcc$, the negative log-likelihood is 
displayed in Fig.~\ref{fig:mixing}a as a function of $\cos\theta_{\rm mix}$. 
Not surprisingly, the Z peak data (Section~\ref{sec:lepslc}) 
constrain the coupling of the sbottom to the Z to be vanishingly 
small, $\cos\theta_{\rm mix} =0.39 \pm 0.07$. 
\eject
The value of the mixing angle was therefore fixed to $\cos\theta_{\rm mix} 
=0.39$. The combined negative log-likelihood and 
those for PEP/PETRA, TRISTAN and LEP\,2 data are shown in 
Fig.~\ref{fig:mixing}b as a function of $\alpha$. (For LEP\,1 and SLC, 
the likelihood does not depend on $\alpha$, because of the vanishing 
sbottom cross section for $\cos\theta_{\rm mix} =0.39$.) 
The values of $\alpha$ for which the different negative log-likelihood 
functions are minimized are indicated in Table~\ref{tab:alpha}, 
together with the corresponding 68\% confidence intervals and the 
95\%\,C.L. upper limits on $\alpha$. (This one-sided upper limit is the
$\alpha$ value for which the negative log-likelihood increases by 
$1.64^2/2$ with respect to the minimum.)
\begin{table}[htbp]
\begin{center}
\caption{\footnotesize The values $\alpha_{\rm min}$ for which the negative 
log-likelihood is minimized in PEP/PETRA, TRISTAN and LEP\,2 data, and in 
the combination, together with the 68\% confidence intervals and the 
95\%\,C.L. upper limits, $\alpha_{95}$, for $\cos\theta_{\rm mix} =0.39$
and $\msbot=6\,\gevcc$.
\label{tab:alpha}}
\vspace{3mm}
\begin{tabular}{|c|c|c|} \hline\hline
Data & $\alpha_{\rm min}$ & $\alpha_{95}$  \\ 
\hline\hline
  PEP/PETRA  & $0.45\pm0.30$ & 0.94 \\
  TRISTAN    & $0.21\pm0.39$ & 0.85 \\
  LEP\,2     & $1.32\pm0.74$ & 2.52 \\
  All        & $0.45\pm0.23$ & 0.82 \\
\hline\hline
\end{tabular}
\end{center}
\end{table}

As was already alluded to in Section~\ref{sec:lowdata}, the lower energy data 
do not favour the sbottom hypothesis ($\alpha=1$). They are, instead, 
compatible with the standard model ($\alpha=0$) within one standard 
deviation or thereabout. 
A slight excess in the LEP\,2 data, at the $1.7\sigma$ level 
(Section~\ref{sec:lep2}), translates as such to the combined result. 
The latter, however, excludes the sbottom hypothesis
with $\msbot\ = 6\,\gevcc$ at more than 95\%\,C.L., when no systematic 
uncertainty is assigned to the standard model prediction.

The main sources of uncertainty for the theoretical prediction
of the $\epemto\ \qqbar$ cross section are {\it (i)} the knowledge and 
the running of the strong coupling constant $\alpha_S$; {\it (ii)} the 
running of the electromagnetic coupling constant $\alpha_{\rm QED}$; and 
{\it (iii)} the theoretical accuracy of the prediction from the 
{\tt ZFITTER} program. As in Ref.~\cite{janot}, the values and the 
uncertainties of the strong and electromagnetic coupling constants 
were taken to be
\begin{equation} 
\alpha_S(\mZ) =  0.1183 \pm 0.0020~\cite{bethke} \ \ \ {\rm and} \ \ \ 
\alpha(\mZ)^{-1} = 128.95 \pm 0.05~\cite{lep}, \ \ \ 
\end{equation}
leading to uncertainties in the hadronic cross section prediction of
0.15\% and 0.08\%, respectively. The missing higher orders in {\tt ZFITTER}
are estimated to contribute another 0.1\%. These numbers add quadratically
to a total systematic uncertainty $\eta_{\rm th}$ of the order of 0.2\%, 
in agreement with the estimate of Ref.~\cite{lep} ($\eta_{\rm th} = 0.26\%$) 
for LEP\,2 data. 

If this common systematic uncertainty is assumed to have a Gaussian
probability density function, the negative log-likelihood can be modified 
as follows, to account for the full correlation between all 
centre-of-mass energies:
\begin{equation}
{\cal L} = {1\over 2}\sum_{i,j=1}^N \Delta_i^\prime S_{ij}^{-1} 
\Delta_j^\prime 
+ {\rho_{\rm th}^2 \over 2 \eta_{\rm th}^2}
{\rm \ \ with \ \ }
\Delta_i^\prime = \sigma_{{\rm had},i} - 
\left[ (1+\rho_{\rm th})\sigma_{{\rm th},i} + \alpha 
\sigma_{{\rm NP},i} \right],
\end{equation}
where $\rho_{\rm th}$ is the actual theoritical bias of the standard model 
prediction, to be fitted from the data. 

\begin{figure}[htbp]
\begin{picture}(160,105)
\put(20,-7){\epsfxsize120mm\epsfbox{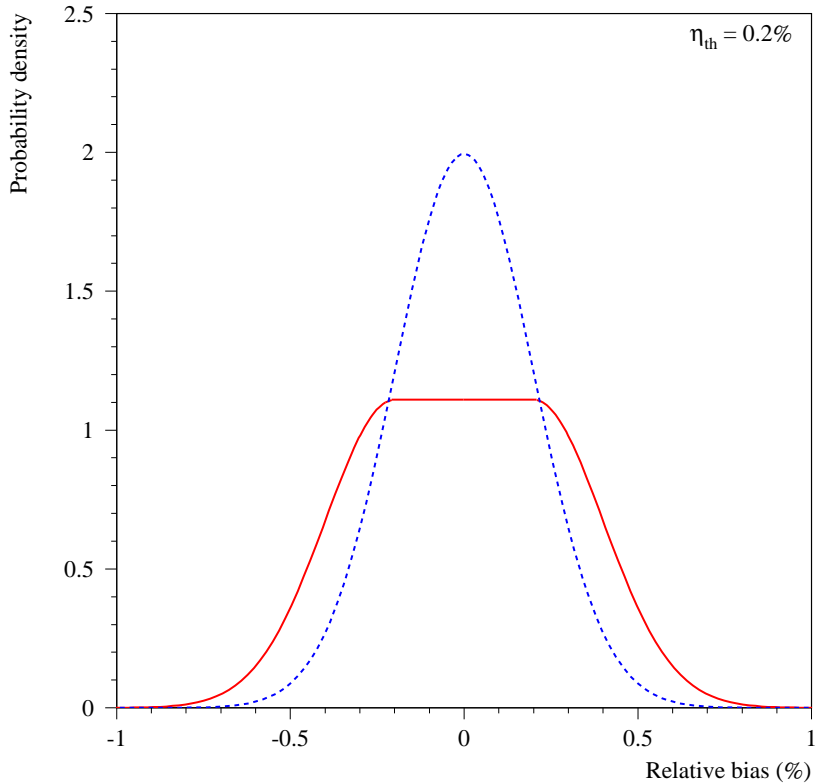}}
\end{picture}
\caption[ ]
{\protect\footnotesize Probability density function for the 
conventional (Gaussian) systematic uncertainty treatment (dashed 
curve) and suggested here instead (full curve) to account for the 
non-Gaussian nature of theory uncertainties, with $\eta_{\rm th}=0.2\%$. 
\label{fig:ansatz}}
\end{figure}

It is reasonable, however, to take into account the non-Gaussian nature of 
uncertainties of theoretical origin. 
For example, the missing higher orders in {\tt ZFITTER} may turn into a 
bias of $-0.1\%$, $0\%$ or $0.1\%$ with an equal probability. (In fact, 
the least likely value is certainly 0\%, as missing orders are expected 
to contribute a finite amount to the cross section.) Similarly, the 
uncertainty on the absolute value of $\alpha_S(\mZ)$ is dominated by 
theory, and cannot be considered as Gaussian. It is therefore probably 
more adequate to assume a probability density function as displayed in 
Fig.~\ref{fig:ansatz}, {\it i.e.}, flat between $-\eta_{\rm th}$ and 
$+\eta_{\rm th}$, and with a Gaussian shape outside this interval. 

\noindent
The likelihood was therefore further modified by changing the
$\rho_{\rm th}^2/2\eta_{\rm th}^2$ term to
\begin{eqnarray}
(\rho_{\rm th}+\eta_{\rm th})^2/2\eta_{\rm th}^2 & 
{\rm if\ } &  \phantom{-\eta_{\rm th} < } \rho_{\rm th} < -\eta_{\rm th},\\
0 \phantom{2\eta_{\rm th}^2} & 
{\rm if\ } & -\eta_{\rm th} < \rho_{\rm th} < \eta_{\rm th},\\
(\rho_{\rm th}-\eta_{\rm th})^2/2\eta_{\rm th}^2 & 
{\rm if\ } & \phantom{-\eta_{\rm th} < } \rho_{\rm th} > \eta_{\rm th}.
\end{eqnarray}

\begin{figure}[h]
\begin{picture}(160,110)
\put(15,-8){\epsfxsize130mm\epsfbox{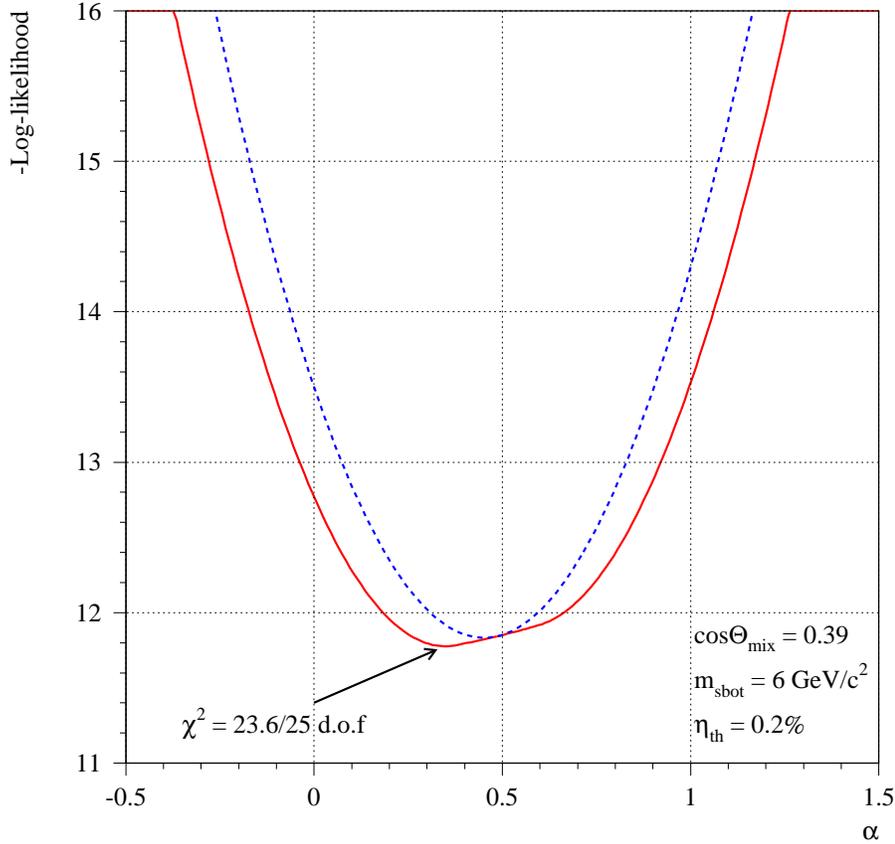}}
\end{picture}
\caption[ ]
{\protect\footnotesize The combined negative log-likelihood curves with 
theoretical systematic uncertainties included, assumed to be Gaussian 
(dashed curve) or non-Gaussian (full curve), as a function of $\alpha$ 
for $\cos\theta_{\rm mix} = 0.39$ and $\msbot\ = 6\,\gevcc$.
\label{fig:final}}
\end{figure}

This negative log-likelihood was then minimized with respect to the
theoretical bias $\rho_{\rm th}$, for each value of $\alpha$, with Gaussian 
and non-Gaussian uncertainties. The result is displayed in 
Fig.~\ref{fig:final} in the two configurations as a function of $\alpha$, 
for $\cos\theta_{\rm mix} = 0.39$ and $\msbot\ = 6\,\gevcc$. The values of 
$\alpha$ for which the negative log-likelihood is minimized are indicated 
in Table~\ref{tab:final}, together with the corresponding 68\% confidence 
intervals and the 95\%\,C.L. upper limits on $\alpha$. 

\begin{table}[t]
\begin{center}
\caption{\footnotesize The values $\alpha_{\rm min}$ for which the combined 
negative log-likelihood is minimized for Gaussian and non-Gaussian 
uncertainties, together with the 68\% confidence intervals and the 
95\%\,C.L. upper limits, $\alpha_{95}$, for $\cos\theta_{\rm mix} =0.39$
and $\msbot=6\,\gevcc$. The fit results for PEP/PETRA, TRISTAN and LEP\,2
($\alpha_{\rm PETRA}$, $\alpha_{\rm TRISTAN}$ and $\alpha_{\rm LEP\,2}$) 
are also given.
\label{tab:final}}
\vspace{5mm}
\begin{tabular}{|c|c|c|c|c|c|} \hline\hline
%\rule{0pt}{4mm}
Uncertainties & $\alpha_{\rm min}$ & $\alpha_{95}$ 
& $\alpha_{\rm PETRA}$ & $\alpha_{\rm TRISTAN}$ & $\alpha_{\rm LEP\,2}$ \\ 
\hline\hline
%\rule{0pt}{4mm}
Gaussian & $0.45 \pm 0.24$ & 0.85 & $0.45\pm 0.35$ & $0.16\pm 0.47$ & $1.68 \pm 1.02$ \\
\rule{0pt}{4mm}
  Non-Gaussian & $0.34^{+0.42}_{-0.24}$   & 0.92 
&  $0.59^{+0.31}_{-0.57}$ &  $0.06^{+0.69}_{-0.48}$ & $1.87 \pm 1.02$ \\
\hline\hline
\end{tabular}
\end{center}
\end{table}
It can be seen that 
the upper limit on $\alpha$ depends very little on the way the common 
systematic uncertainties are dealt with. The most conservative approach 
is chosen here to derive the final results.
\null\vfill
\vfill\eject
\section{Results}
\label{sec:results}

\begin{figure}[htbp]
\begin{picture}(160,110)
\put(15,-8){\epsfxsize130mm\epsfbox{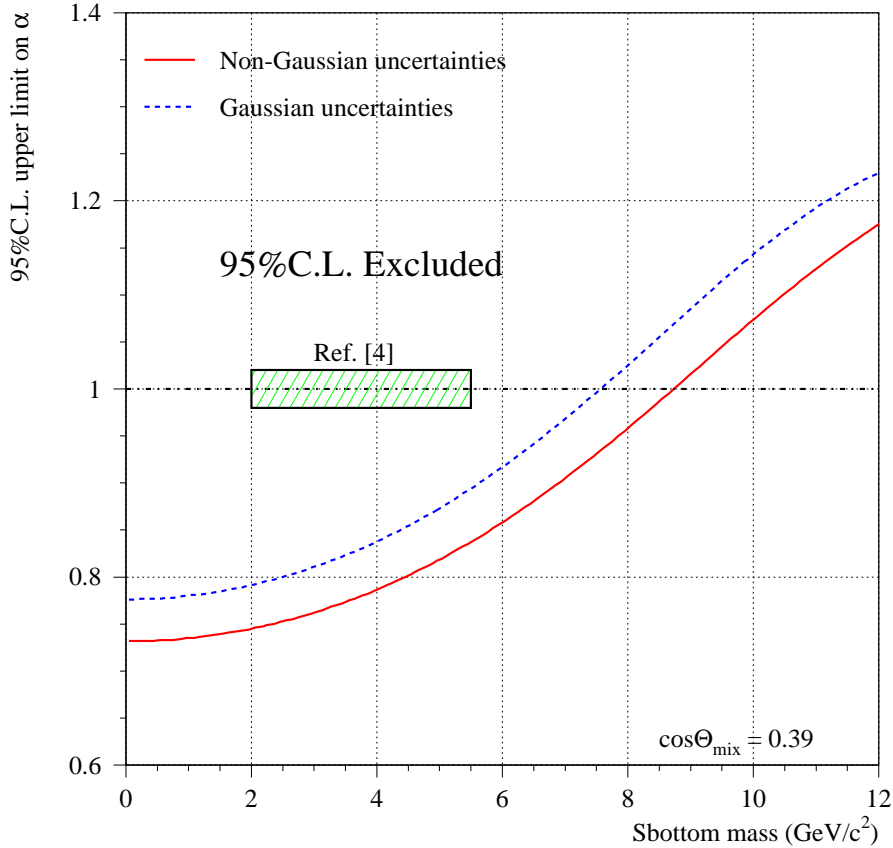}}
\end{picture}
\caption[ ]
{\protect\footnotesize The 95\%\,C.L. upper limit on $\alpha$ as a 
function of the sbottom mass, with non-Gaussian (full curve) and Gaussian 
(dashed curve) common systematic uncertainties. Also shown are the 
predictions of the model of Ref.~\cite{berger}, now excluded by this 
analysis.
\label{fig:limits}}
\end{figure}

The same procedure was repeated by varying the sbottom mass from 0 to 
12\,\gevcc. For each mass, the 95\%\,C.L. upper limit on $\alpha$
was determined as explained above. A sbottom with a given mass is excluded 
if this upper limit is smaller than unity. Figure~\ref{fig:limits} shows 
the 95\%\,C.L. upper limit on $\alpha$ for $\cos\theta_{\rm mix} = 0.39$ 
as a function of the sbottom mass, with Gaussian and non-Gaussian 
uncertainties. (In the latter configuration, the non-Gaussian nature 
of the likelihood was taken into account in the determination of the 
limit.) Sbottom masses below 7.5\,\gevcc\ are excluded at the 95\% 
confidence level.

\begin{figure}[t]
\begin{picture}(160,110)
\put(15,-8){\epsfxsize130mm\epsfbox{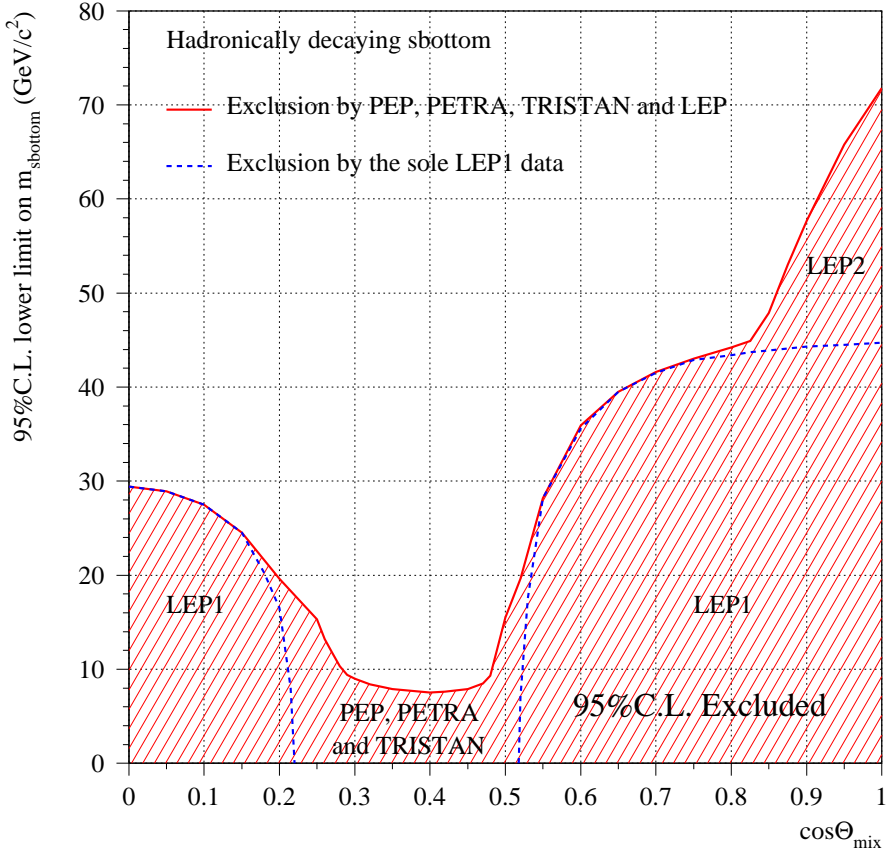}}
\end{picture}
\caption[ ]
{\protect\footnotesize Absolute 95\%\,C.L. lower limit on $\msbot$ as a 
function of $\cos\theta_{\rm mix}$, for hadronically decaying sbottoms. 
The hatched area is excluded at 95\%\,C.L. The dashed line shows the 
exclusion achieved with the sole Z peak data. 
\label{fig:sbottom}}
\end{figure}

Because $\cos\theta_{\rm mix}$ is very much constrained by the Z peak data, 
the upper limit on $\alpha$ is expected to be smaller than that shown in 
Fig.~\ref{fig:limits} for any other value of the mixing angle. As a check, 
the procedure was repeated again by varying $\cos\theta_{\rm mix}$ 
from 0 to 1, with non-Gaussian uncertainties. The resulting sbottom mass 
lower limit is shown in Fig.~\ref{fig:sbottom} as a function of 
$\cos\theta_{\rm mix}$, and is indeed at least 7.5\,\gevcc\ over the whole 
range. (The region excluded by LEP\,2 data at large values of  
$\cos\theta_{\rm mix}$ is probably over-optimistic, as four-jet events -- 
expected from such heavy sbottom pair as well as W pair production -- are 
rejected from the \qqbar\ event samples selected above the WW threshold.) 

It is worth mentioning that the presence of a light sbottom would 
slow down the running of $\alpha_S$ with the centre-of-mass
energy. (It would be even more so with an additional light gluino.)
Starting from the value accurately measured in $\tau$ decays~\cite{tau}, 
(the only measurement not affected by a sbottom heavier than 2\,\gevcc\
and lighter than 5.5\,\gevcc, and corresponding to 
$\alpha_S(\mZ)=0.121\pm0.003$ in the standard model), this slower running 
would lead to values of $\alpha_S$ larger than assumed in this letter, 
at all centre-of-mass energies. The total New Physics contribution 
(from the direct sbottom production and the increase of $\alpha_S$) 
would further increase the effect on the total hadronic cross section 
expected at PEP, PETRA, TRISTAN, SLC and LEP. The 7.5\,\gevcc\ lower 
limit on the sbottom mass is therefore probably very conservative.

\section{Conclusion}
\label{sec:conclusion}

The \epemto\ hadron cross section data collected well above the \bbbar\
resonances have been compiled and analysed to search for an anomalous
production of hadronic events. Altogether, the PEP, PETRA, TRISTAN, 
LEP\,1, SLC and LEP\,2 data allow a light sbottom decaying hadronically 
to be excluded at 95\%\,C.L. for any mixing angle, if its mass is below 
7.5\,\gevcc. When combined with the result of Ref.~\cite{aafke} in which 
a stable sbottom with mass below 92\,\gevcc\ is excluded, this 
analysis definitely invalidates the model of Ref.~\cite{berger}
with a 12-16\,\gevcc\ gluino and a 2-5.5\,\gevcc\ sbottom.

\subsection*{Acknowledgments}

I am grateful to German Rodrigo, Sabine Kraml and the 
CERN Journal Club on Phenomenology for having suggested this 
analysis to me, and to G\"unther Dissertori, Vladimir Ezhela,
Michelangelo Mangano, Ramon Miquel, Michael Schmitt, Daniel Treille
and Oleg Zenin  for interesting discussions. Many thanks are due to 
Bolek Pietrzyk, Helmut Burkhardt, Gerardo Ganis and Brigitte Bloch 
for their help with {\tt ZFITTER} and other Monte Carlo programs, 
and for many enlightening exchanges. Finally, I am indebted to my 
Physics Letters referees for many contructive suggestions.

%Above all, I would like to warmly thank Michael Schmitt for his accurate
%remarks and his collaborative help in the analysis of the important
%drawbacks affecting the public computer files of the hadronic cross
%section data.

{\small
\subsection*{\it Note added}
This work has been primarily motivated by the ``apparent excess'' reported
in Ref.~\cite{schmitt}. With the collaborative help of the author, the excess
was found to be an artifact of duplicated, missing and over-corrected data
in the computer-readable files of the Review of Particle 
Physics~\cite{zenin,pdgfiles} augmented by an incorrect interpretation 
of the Z peak data in Ref.~\cite{schmitt}. The aforementioned 
computer-readable files are being updated to include the work described 
in this letter.
}

\end{document}